# Phenomenological Model of the Nonlinear Microwave Response of a Superconductor Containing Weak Links

Anton V. Velichko[1,2] and Adrian Porch[1],
[1]School of Electronic & Electrical Engineering, University of Birmingham, UK
[2]Institute of Radiophysics & Electronics of NAS, Ukraine

A phenomenological model is proposed which describes the effect of both dc and rf magnetic fields, $H$, on the microwave surface impedance, $Z_s = R_s + jX_s$, of a superconductor containing weak links. Two types of the weak links are considered; a weak link between two grains, shunted by another grain, and a non-shunted weak link. In both cases, the dependences of $R_s$ and $X_s$ on applied $H$ were found to be anomalous. Under certain conditions, the values of $Z_s(H)$ can fall below the zero-field values. Comparison with experiment is performed, and very good qualitative and (in some cases) quantitative agreement is found.

## I. INTRODUCTION

Microwave breakdown, being one of the most fascinating phenomenon of the contemporary physics of superconductivity, is not yet fully understood. A few simple phenomenological models[1–3] developed recently have helped to shed some light on the mechanisms of the microwave nonlinearity in high-temperature superconductors (HTS). However, in the last few years, there have appeared some observations of quite unusual features in both the rf and dc field dependences of the microwave surface impedance, $Z_s = R_s + jX_s$. Both the surface resistance, $R_s$, and the surface reactance, $X_s$, were reported to change non-monotonically with field in epitaxial HTS thin films.[4–8] Here, the values of $R_s$ and $X_s$ in applied fields, $H$, were sometimes seen to fall below the relevant zero-field values. Several hypotheses have been put forth to explain these peculiarities in $Z_s(H)$. Among them are stimulation of superconductivity[6,7,9], and field-induced alignment of magnetic impurity spins.[4,7,10] However, until now no theoretical model of $Z_s(H)$ incorporating the above mentioned scenarios has been proposed.

An anomalous behavior in $Z_s(H)$ is not a new feature solely inherent to HTS only. Similar phenomena have also been reported for low-$T_c$ superconductors (see, e. g., Ref.[11,12]) subject to simultaneously applied ac and dc magnetic fields. In particular, low-frequency anomalies in $Z_s(H)$ have been ascribed to the field-dependent surface energy barrier[12,13]. Though these low-$T_c$ mechanisms may bear something in common with the anomalous behavior reported for HTS, there seem to be other unique properties of the metal oxides, such as intrinsically formed magnetic impurities[10], as well as the enhanced anisotropy and inhomogeneity of HTS, which are currently believed to be responsible for the observed anomalies in $Z_s(H)$[4,5,7].

Recently, a phenomenological model developed by Gallop et al.[14] to explain the difference in the effect of dc and rf magnetic fields upon $Z_s$, was unexpectedly found to give a reduction of the microwave losses ($R_s$) with increased field. In that model, the microwave response of a Josephson weak link (WL) shunted by a superconducting grain was considered. However, no reduction of $X_s$ with the field can be obtained within the framework of Gallop's model, saying nothing about the correlating decrease of both $R_s(H)$ and $X_s(H)$, as was experimentally observed by some groups[4,8].

In the present paper, we present a unified model for the microwave response of a superconductor containing WLs, considered as long Josephson junctions (JJs). We treat the superconductor as an effective medium consisting of two subsystems, the grains and the weak links. Here, the grains possess the properties of the bulk superconductor, and the weak links may be either a superconductor with reduced critical parameters ($H_{c1}$, $J_c$ etc.) or a normal conductor with induced superconductivity via the proximity effect. The surrounding grains and the WL itself are modelled by an equivalent circuit, the elements of which, resistivities and inductivities, represent dissipative and inertial processes in the material. Two types of WLs are considered. Firstly, a simple WL between two grains. Secondly, a WL shunted by another third grain[15], which is strongly coupled to the other two grains. These two topologies are hereafter referred to as "the non-shunted" and "the shunted WL geometry", respectively. The second geometry was previously considered by Gallop et al.[14], using the effective medium approximation[16] for expressing the Josephson penetration depth, $\lambda_J$ as a function of the average grain size, $a$. The above approach, however, is justified only if both $\lambda_J$ and $\lambda_L \gg a$, where $\lambda_L$ is the London penetration depth.

Another main assumption of this paper (in addition to the concept of the shunting grain) is that for high-quality HTS films, due to high critical current densities, $J_c$ (in excess of $10^{10}$ A/m$^2$), and larger grain sizes (up to several $\mu$m, see, e. g., Ref.[17]), the effective medium approximation for $\lambda_J$ breaks down and the weak links have to be considered as long JJs. In this case, $\lambda_J$ is no longer dependent on $a$, but rather depends on the magnetic thickness of the junction, $d_m = 2\lambda_L + d$; here, $d$ is the grain separation. Another important feature of the model presented here is that the junction resistivity, $\rho_J$ is considered to be magnetic field-dependent due to modulation of the quasiparticle transport at low fields and flux flow dissipation at higher fields. This field dependence, which was always ignored in earlier weakly-coupled grain models (see, e. g., Ref.[18]), proved to be very important and allowed the description of the majority of the unusual features in $Z_s(H)$ of HTS thin films both at qualitative and (in some cases) quantitative levels.

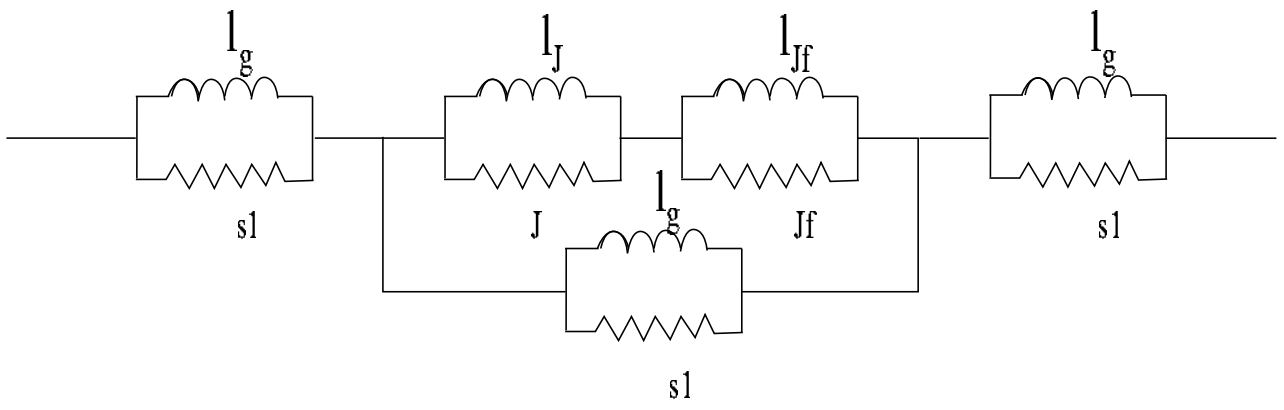

FIG. 1. Equivalent circuit of a grain-shunted weak link. Here $\rho_{Jf}$ and $l_{Jf}$ are real and imaginary parts of the complex flux flow resistivity, with $\tilde{\rho}_{Jf}$ given by Eq.(11). In the case of the non-shunted weak link, the lowermost part of the circuit representing the shunting grain is absent.

## II. THE MODEL

The model presented here is based on the equivalent circuit description of superconductors.[19] The general equivalent circuit, comprising both the situations modelled in the present paper, is shown in Fig. 1. The effective complex resistivity of the grains, according to Fig. 1, in the limit of $\rho_{s1} \gg \omega l_s$ (which is always valid for frequencies $f \leq 100$ GHz ($f = 2\pi/\omega$) and temperatures $T$ not too close to the critical temperature, $T_c$), can be written as

$$\tilde{\rho_g} \simeq \frac{(\omega l_s)^2}{\rho_{s1}} + j\omega l_s. \qquad (1)$$

Note that in the above limit, $l_g = l_s$; here, $l_s$ and $l_g$ are the effective and the real grain inductivities, respectively.

The main assumptions of the model are as follows. The magnetic field (rf and dc) affects the critical current density of the junction, $J_c$ and the junction resistivity, $\rho_J$, as well as the bulk (grain) penetration depth, $\lambda_g$. Both rf and dc current densities, $J_{rf}$ and $J_{dc}$, associated with applied rf and dc fields, are assumed to be small enough, so that not to cause the flux penetration into the grains. In order words, within the whole range of the applied field, grains are assumed to be in the Meissner state. However, nonlinear dissipation arises within the grains due to pair breaking produced by the rf current. According to the theory of Dahm and Scalapino[20], for a superconductor with $d_{x^2-y^2}$-wave symmetry of the order parameter, the superfluid density, $n_s$ should vary with magnetic field as follows

$$\frac{n_s(T,H)}{n} = \left(\frac{\lambda_g(0)}{\lambda_g(T)}\right)^2 \cdot \left[1 - b_\theta(T)(H/H_g)^2\right], \qquad (2)$$

where the coefficient $b_\theta$ depends on $T$ and the direction of current flow with respect to the planar Cu-O bonds (or antinodes of the gap function, $\Delta(\theta, T) = \Delta_0(T)\cos(2\theta)$). The scaling field, $H_g$ is of the order of the thermodynamic critical field, $H_c$. From Eq.(2) it follows that the nonlinear contribution to the penetration depth and quasiparticle resistivity can be written as

$$\lambda_g^2(T,H) = \lambda_g^2(T) \cdot \left[1 + b_\theta(T)(H/H_g)^2\right], \qquad (3)$$

and

$$\rho_{s1}(T,H) = \frac{m}{e^2\tau\big(n - n_s(T,H)\big)}, \qquad (4)$$

where $m$, $e$ and $\tau$ are mass, charge and scattering time of the quasiparticles, and $n$ is the total electron density.

From Eq.(1) it follows that the real part of $\tilde{\rho}_g$ is given by

$$\rho_g(H) = \frac{(\omega l_g(H))^2}{\rho_{s1}(H)}, \qquad (5)$$

where $\rho_{s1}$ and $l_g$ are field-dependent quantities in accordance with Eq.(3) and Eq.(4).

The field dependence of the critical current of a Josephson tunnel junction is given by[21]

$$I_c(H) = I_{c0} \cdot \frac{\sin(\pi\Phi/\Phi_0)}{\pi\Phi/\Phi_0} \sim I_{c0} \cdot \frac{\sin(H/H_J)}{H/H_J}, \qquad (6)$$

where $H_J$ is the relevant scaling field ($\sim H_{c1J}$, the junction lower critical field). However, according to Mahel et al.[22], for an ensemble of JJs with a random distribution of critical currents and dimensions, the following magnetic field dependence of $I_c$ is expected

$$I_c(H) = I_{c0} \cdot \frac{1}{1 + H/H_J}, \qquad (7)$$

where $H_J$ is the junction scaling field, which equals $\Phi_0/(2\mu_0\lambda_J(0)\lambda_g(0))$ for a long junction. The above equation represents the envelope of the Fraunhofer pattern, Eq.(6), at high fields. Moreover, this type of $I_c(H)$ is often observed experimentally (see, e. g., Ref.[17] and Ref.[19]), and therefore, we adopt it in our model.

Since in a long JJ the transport current is not distributed homogeneously, but rather flows only within the penetration depth, $\lambda_J$, the critical current density, $J_c$ is related to $I_c$ as follows (see, e. g., Ref.[23])

$$J_c(H) = I_c(H)/S_{eff} = I_c(H)/[4\lambda_J(H)(a - \lambda_J(H))], \qquad (8)$$

where $S_{eff}$ is the effective area of the junction through which the current flows.

The Josephson penetration depth can be described as (see, e. g., Ref.[21])

$$\lambda_J(H) = \sqrt{\frac{\hbar}{2e\mu_0 J_c(H) d_m}}. \quad (9)$$

We further assume that the junction resistivity, $\rho_J$ is also dependent on magnetic field via two different non-linear mechanisms in the junction, namely non-linear pair breaking (at low $H$) and flux flow (at higher $H$). Similarly to the grain resistivity, $\rho_g$, the expression for the junction resistivity can be written as

$$\rho_J(T,H) = \rho_{J0}\big(n - n_s(T,H)\big), \quad (10)$$

where $\rho_{J0}$ is the low field value of the junction resistivity. We should emphasize here that though the weak link material is in general different from that of the bulk, we consider it to be superconducting with reduced critical parameters. Therefore, we apply the same Drude model to describe the conductivity (resistivity) of the weak links as a function of the quasiparticle concentration. Here we neglect such processes as Josephson and quasiparticle quantum tunneling through the WLs, which might lead to qualitatively different field dependence of the junction resistivity. Such an assumption is justified only if the characteristic size of the WL regions in the *ab* plane is big enough compared to the coherence length, $\xi$ ($\sim$ 2 nm for YBaCuO at low $T$). Interestingly, recent atom-probe field ion microscopy experiments by Hu et al.[24] have demonstrated that in non-optimally doped YBaCuO, the width of an oxygen deficient region around the twin boundaries (which are the main type of weak links in YBaCuO films) can be as large as 6–7 nm. This observations suggest that the direct quantum (Josephson or quasiparticle) tunneling may be not the main mechanism of the charge transfer through the natural WLs in HTS materials.

Another contribution to the junction resistivity at high fields in our model comes from the flux flow mechanism[25]

$$\tilde{\rho}_{Jf}(H) = \frac{\Phi_0 \mu_0 H}{\eta_J\left[1 + j\ \omega_{Jd}(H)/\omega\right]}, \quad (11)$$

where the depinning frequency, $\omega_{Jd}(H)$ $[\omega_{Jd} = 2\pi f_{Jd}]$ bears the field dependence of the critical current, i. e. $\omega_{Jd}(H)/\omega_{Jd}(0) = J_c(H)/J_c(0)$ (see Eq.(8)). Here, $\eta_J$ is the flux viscosity within the WLs. The inductivities of the grains and the junction are given by

$$l_g = \mu_0\lambda_g^2, \;\; l_J = \mu_0\lambda_J^2, \quad (12)$$

respectively, and the total effective resistivity of the junction is

$$\rho_{eff} = (1-\mu)\tilde{\rho}_g + \mu\left[\frac{1}{\tilde{\rho}_g} + \Big(\ \tilde{\rho}_J + \ \tilde{\rho}_{Jf}\Big)^{-1}\right]^{-1}, \quad (13)$$

where $\tilde{\rho_g}$ is given by Eq.(1), and $\tilde{\rho}_J = (1/\rho_J + 1/j\omega l_J)^{-1}$. Here $\mu$ represents a relative contribution of WLs compared to that of the grains (analogous to the areal WL ratio $\mu$ introduced by Halbritter[17] to describe the surface impedance of granular superconductors). This expression is valid for the shunted WL geometry. For the case of the non-shunted junction, the first term in the square brackets of Eq.(13) has to be omitted. In addition, for the shunted WL geometry, a part of the transport current (both rf and dc) will flow through the shunting grain. If one takes into account that rf circuits usually operate in the current source regime (i. e., $I_{rf}$ =const; see, e. g., Ref.[21]), additional weighting factors have to be introduced in the expressions for the grain penetration depth, Eq.(3), the grain resistivity, Eq.(5), the junction critical current, Eq.(7), and the junction resistivity, Eq.(10) and Eq.(11). These weighting factors should essentially be the ratio of (the real parts of) the relevant conductivities

$$\alpha_J = Re\left[\frac{\Big(\ \tilde{\rho}_J + \ \tilde{\rho}_{Jf}\Big)^{-1}}{\Big(\ \tilde{\rho}_J + \ \tilde{\rho}_{Jf}\Big)^{-1} + \Big(\rho_g + j\omega l_g\Big)^{-1}}\right],$$

$$\alpha_J = 1 - \alpha_g \quad (14)$$

where the indices "g" and "J" denote grains and junction, respectively. For dc currents, these factors will be different, but here we shall not consider the case of a dc field for the shunted junction geometry.
Finally, the surface impedance of the whole structure is

$$Z_s^{eff} = \sqrt{j\omega\mu_0\rho_{eff}}. \quad (15)$$

All the simulations in this paper were performed numerically by simultaneous solving the system of equations (3), (4), (7), (8), (9), (10), (11) and (14).

## III. SAMPLES

The two YBaCuO thin films studied here are deposited by e-beam co-evaporation onto polished (001)-orientated MgO single crystal substrates 10 × 10 mm$^2$. The films are 350 nm thick. The c-axis misalignment of the films is typically less than 1%, and the *dc* critical current density $J_c$ at 77 K is around $2\cdot10^6$ A/cm$^2$. Critical temperature determined by magnetic susceptibility measurements was $\sim$ 87 K for both the samples. The values of $R_s$ are 35 and 50 $\mu\Omega$, and $\lambda$ is 210 and 135 nm, as determined by low-power microwave measurements using the coplanar resonator technique at 8 GHz[26]. Average surface roughness was in the range of 2-4 nm for both of the samples. The characteristics of these films are comparable with those of the best films fabricated at present (for review see, e. g., Ref.[5]), which testifies about very high quality of the samples studied in this work.

## IV. EXPERIMENTAL RESULTS

In the geometry of the shunted weak link only the case of the rf-field dependence of $Z_s$ is considered, with weighting factors for rf-current given by Eq.(14). The only variable parameter in this case is the junction critical current, $J_c(0)$. A remarkable feature of this geometry is that at high $J_c(0)$ ($\geq 5\cdot10^{10}$ A/m$^2$), $R_s(H)$ may decrease monotonously starting from the lowest fields, being accompanied by a sublinear ($\sim H^n, n<1$) increase of $X_s(H)$ [Fig. 2 (a), 2 (b)]. At lower $J_c(0)$, a sharp maximum in $R_s(H)$ appears, which becomes broader and shallower, and, finally, disappears with further decrease in $J_c(0)$ (see dashed and solid lines in Fig. 2 (a)). Eventually, $R_s(H)$ translates into a sublinear function of $H$, similar to $X_s(H)$. It should be noted that with the change in $J_c(0)$, rather insignificant transformation occurs in $X_s(H)$, which almost preserves its functional form, whereas $R_s(H)$ changes dramatically.

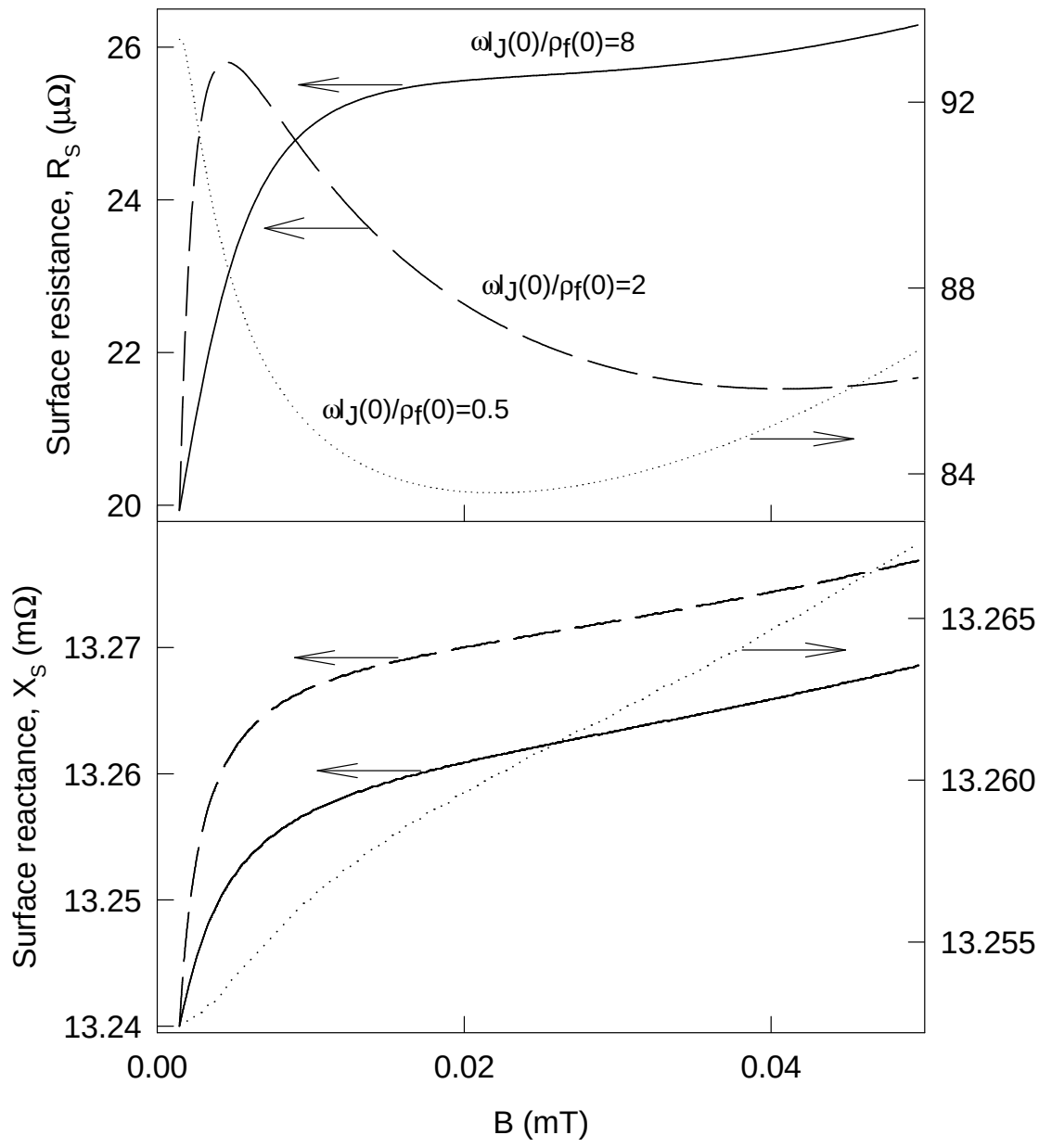

FIG. 2. Simulated magnetic field, $B$ dependence of the surface resistance, $R_s$, and the surface reactance, $X_s$ for the shunted weak link for different values of $J_c(0)$ ($4\cdot10^9$, $1\cdot10^{10}$ and $5\cdot10^{10}$ A/m$^2$ for dotted, dashed and solid lines, respectively). The data shown by dashed lines are offset by 11.5 $\mu\Omega$ ($R_s(B)$) and 7.7 m$\Omega$ ($X_s(B)$) for clarity. The ratio $\omega l_J(0)/\Re[\tilde{\rho}_{Jf}]$, which in this case is the relevant parameter determining the shape of $Z_s(B)$, is given in the figure next to each curve. Other parameters of the simulation are as follows: $f = 8\cdot10^9$ Hz; $\rho_{s1}(0) = 1\cdot10^{-6}$ $\Omega$·m; $B_g = \mu_0 H_g = 250$ mT; $a = 1.0\cdot10^{-6}$ m; $\lambda_g(0) = 2.1\cdot10^{-7}$ m; $\mu = 0.01$; $b_\theta = 1$; $\rho_{J0} = 1.1\cdot10^{-6}$ $\Omega$·m; $\eta_J = 2.0\cdot10^{-10}$ N·s/m$^2$; $f_{Jd0} = 5\cdot10^9$ Hz.

Fig. 3 demonstrates a fitting of experimental data on one of the films studied in this work to the present model at low ($\sim$ 20 K, the main figures) and high ($\sim$ 60 K, the insets) temperatures. Excellent qualitative and quantitative agreements are observed. As one can see, in accordance with the model, with increased $T$ (decreased $J_c$) the shape of $R_s(H)$ undergoes a noticeable change, whereas the functional form of $X_s(H)$ remains nearly the same.

Recently, several research groups (see, e. g., Refs.[4,5,8]), have found a correlated decrease of $R_s(H)$ and $X_s(H)$ in high-quality YBaCuO thin films. Such a behavior can also be reproduced in our model for the non-shunted WL geometry in the limit of relatively small $J_c$ ($\leq 10^8$ A/m$^2$), as shown in Fig. 4. The simulation is performed for three different values of parameter $b_\theta$ [see Eq.(2) and (3)]. As one can see, the increased value of $b_\theta$, which for $d$-wave superconductor can occur at low ($t \leq 0.1$, where $t = T/T_c$) and high ($t \geq 0.7$) temperatures[20], leads to an appearance of a minimum in $R_s(H)$ and $X_s(H)$. Here, the minimum in $X_s(H)$ develops earlier (at lower fields), and is followed by an increase of $X_s$ at higher $H$. This is in a very good qualitative agreement with the experimental results on another film we studied in this work [see Fig. 2 (a), 2 (b)], though we didn't perform fitting to the experimental data in this case.

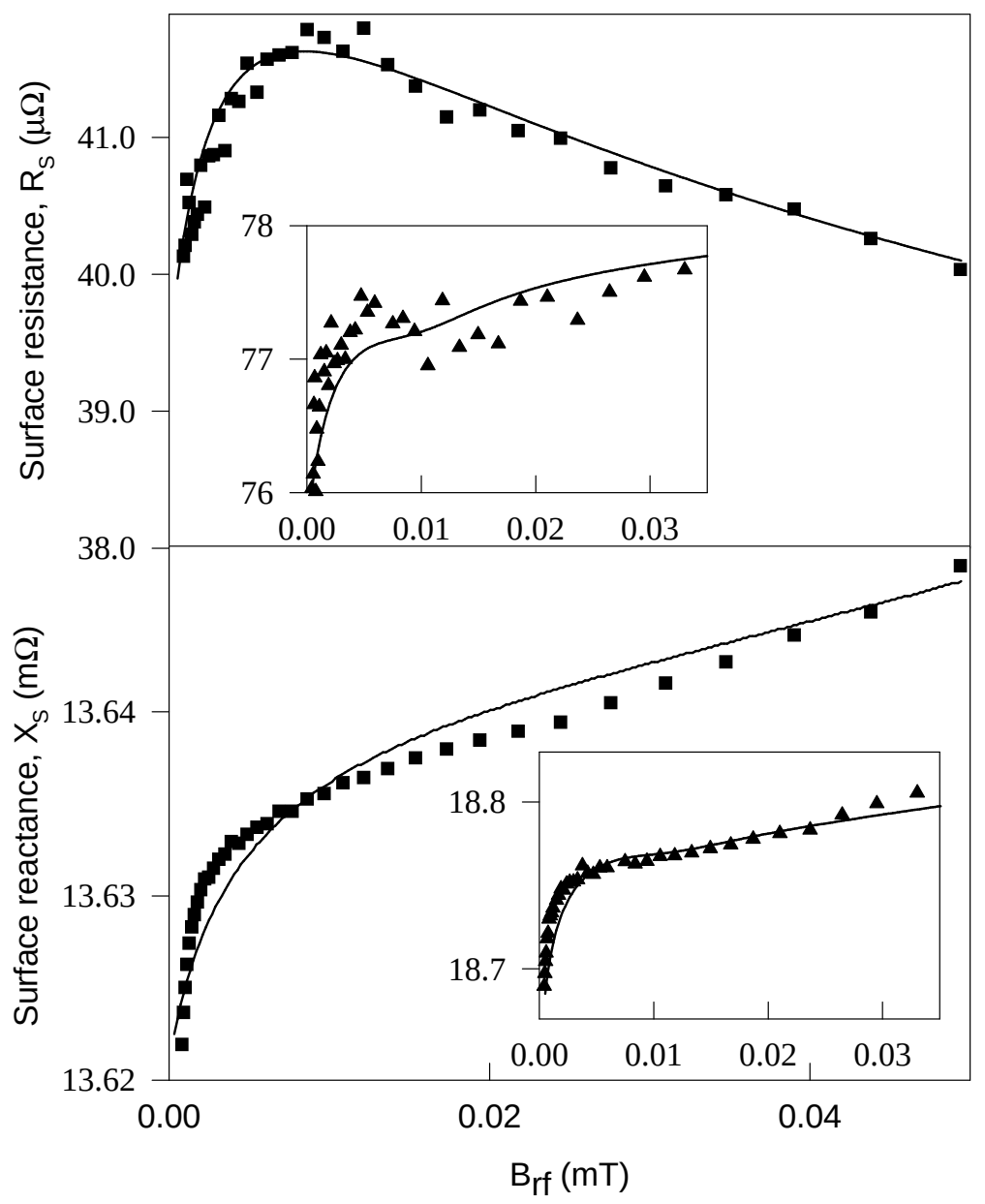

FIG. 3. Experimental data on $R_s(B_{rf})$ and $X_s(B_{rf})$ for a high-quality YBaCuO thin film at $T = 20$ K (the main figures) and $T = 60$ K (the insets) together with theoretical curves obtained within the model. The parameters used for the fitting are as follows: $f = 8\cdot10^9$ Hz; $\rho_{s1}(0) = 1\cdot10^{-6}$ $\Omega$·m; $B_g = \mu_0 H_g = 250$ mT; $a = 6.5\cdot10^{-7}$ m; $\lambda_g(0) = 2.13\cdot10^{-7}$ m; $\mu = 0.05$; $b_\theta = 1$ $J_c(0) = 1.9\cdot10^9$ A/m$^2$ $\rho_{J0} = 5.9\cdot10^{-8}$ $\Omega$·m $\eta_J = 1.9\cdot10^{-10}$ N·s/m$^2$; $f_{Jd0} = 5\cdot10^9$ Hz;
b) $\lambda_g(0) = 2.22\cdot10^{-7}$ m; $J_c(0) = 1.2\cdot10^9$ A/m$^2$; $\mu = 0.15$; $\rho_{J0} = 5.2\cdot10^{-7}$ $\Omega$·m. The other parameters are the same as for a).

## V. DISCUSSION AND CONCLUSION

The correlated anomalous decrease of $R_s(H)$ and $X_s(H)$ for the non-shunted geometry takes its origin in the field dependent quasiparticle resistivity of the junction, $\rho_J$, which decreases with $H$. The latter, in turn, originates from the modulation of the quasiparticle density by the magnetic field via the non-linear pair breaking effect. According to Dahm and Scalapino[20], the pair breaking in HTS can be significantly facilitated at low ($t < 0.1$) and high ($t > 0.7$) temperatures due to temperature dependent coefficient $b_\theta$, which may result in a noticeable reduction of the scaling field, $H_g$ as compared to the case of $s$-wave superconductor. For instance, as we found in our simulations, an increase in $b_\theta$ leads to appearance of a characteristic upturn in both $R_s(H)$ and $X_s(H)$, and their deviation from a simple power-law behavior ($R_s, X_s \sim H^n, n < 1$), which we indeed observed in our samples (see experimental data in Fig. 3). However, such an anomalous decrease of both $R_s(H)$ and $X_s(H)$ in the non-shunted WL geometry can only be obtained in the limit of $\omega l_J \gg \rho_J$, which assures that both field-dependences of $R_s$ and $X_s$ of the junction are due to that of the junction resistivity. Bearing in mind that from physical considerations, parameters such as the low power penetration depth, $\lambda_g(0)$ and the grain size, $a$ are allowed to vary within rather narrow intervals ($130 \leq \lambda_g \leq 200$ nm, $0.1 \leq a \leq 10$ $\mu$m), we have to impose quite strict limitations on the value of other parameters, such as $J_c$ and $\rho_{Jn}$. In fact, the above limit can be well satisfied only for rather low $J_c$ ($\lesssim 10^8$ A/m$^2$) and high-conductivity ($\rho_{Jn} \lesssim 10^{-7}$ $\Omega$·m) junctions, which are likely to be SNS-type JJs.

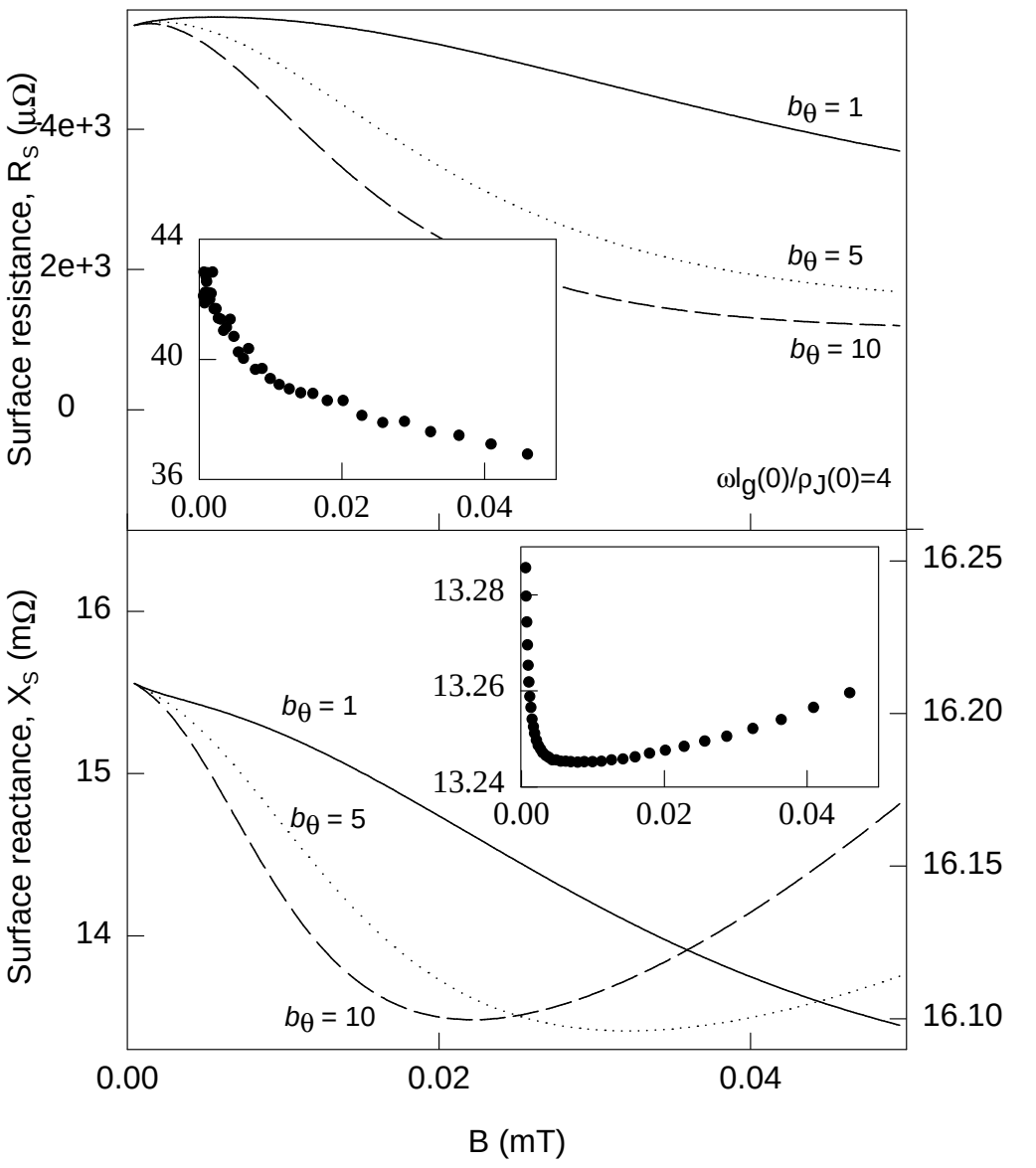

FIG. 4. Simulated results on $R_s(B)$ and $X_s(B)$ for the non-shunted WL geometry in the low-$J_c$ and low-$\rho_{J0}$ limits for various values of $b_\theta$. The ratio $\omega l_J(0)/\rho_J(0)$, which in this case is the relevant parameter determining the behavior of $Z_s(B)$, is given in the figure. Other simulation parameters are as follows: $f = 8 \cdot 10^9$ Hz; $\rho_{s1}(0) = 1 \cdot 10^{-6}$ $\Omega$·m; $B_g = \mu_0 H_g = 250$ mT $a = 2.0 \cdot 10^{-6}$ m; $\lambda_g(0) = 2.1 \cdot 10^{-7}$ m; $\mu = 0.05$; $J_c(0) = 1.7 \cdot 10^8$ A/m$^2$; $\rho_{J0} = 5.8 \cdot 10^{-8}$ $\Omega$·m; $\eta_J = 1.0 \cdot 10^{-8}$ N·s/m$^2$; $f_{Jd0} = 5 \cdot 10^9$ Hz. The insets show experimental data for the other YBaCuO film at 15 K.

It has to be mentioned here, that the implications of the d-wave concept in terms of the pair breaking parameter, $b_\theta$ have not yet been confirmed experimentally. For instance, decrease of the critical current density or thermodynamical critical field, which is theoretically expected at low temperatures (due to the expected increase in the value of $b_\theta$), have not yet been found. However, it is well known (see, e. g., Ref.[4,5,7]) that the microwave anomalies in HTS films are usually observed at low temperatures (typically below 15–20 K). Besides, as one can infer from Fig. 4, our model predicts a more profound decrease in both $R_s(H)$ and $X_s(H)$ with increased $b_\theta$, which is expected at low $T$. The increased value of $b_\theta$ also leads to much better qualitative agreement with the experimental data at low temperature (see the inset in Fig. 4), which might be an indirect indication of the expected $d$-wave scenario.

In principle, a correlated decrease of $R_s(H)$ and $X_s(H)$ can also be obtained in the non-shunted geometry. However, in this case a more strict condition, $\rho_J \leq \omega l_g \ll \omega l_J$, has to be imposed on the junction parameters. For typical values of $\lambda_g \sim 2 \cdot 10^{-7}$ m for HTS films, the latter condition means that the quasiparticle resistivity of the junction, $\rho_J$ has to be $\lesssim 5 \cdot 10^{-9}$ $\Omega$·m, which is comparable to that of the highest conductivity metals (such as Au, Cu or Ag). Though one could imagine an artificial junction of that type, one can hardly expect that such WLs may exist in real HTS materials.

As far as the uncorrelated behavior of $R_s(H)$ and $X_s(H)$ is concerned, which is observed for the shunted WL geometry only, it should be emphasized that this is due to the interplay between the field dependences of the inductive component, $\omega l_J$ and the flux-flow resistivity, $\rho_{Jf}$ of the junction. Though the quasiparticle resistivity, $\rho_J$ is also field dependent, it does not contribute much in this case since the scaling field for $\rho_J(H)$ is $\sim H_{cg}$, which is much higher than appropriate scaling field for $l_J(H)$, which is $\sim H_{c1g}$. The major difference of this geometry as compared to the non-shunted WL is that here we can consider the limit of sufficiently higher critical currents ($J_c \geq 10^9$ A/m$^2$), which makes the inductive contribution, $\omega l_J$ come into play, and which is much more realistic regime for high-quality HTS thin films having $J_c$ in excess of $10^{10}$ A/m$^2$ at 77 K.

The three types of $R_s(H)$ behavior in Fig. 2 can be explained as follows. For high $J_c(0)$-junctions, the limit $\omega l_J \leq \rho_{Jf} \ll \rho_J$ is realised, where the overall behavior is governed by $\rho_{Jf}(H)$. Note that in this case, as opposed to the non-shunted WL, an increase in $\rho_{Jf}(H)$ produces a decrease in $R_s(H)$ (rather than an increase as in the other geometry), since increasing $\rho_{Jf}(H)$ makes more and more current divert through the shunting grain, making the total response less and less dissipative. With decreased $J_c(0)$, the inductive term $\omega l_J$ comes into play, which produces the initial increase in $R_s(H)$ followed by decrease being again due to $\rho_{Jf}(H)$. The result of this competition is a maximum in $R_s(H)$ [dashed curve, Fig. 2 a]. Finally, for low $J_c(0)$-junctions, when the limit $\rho_{Jf} \ll \omega l_J \ll \rho_J$ is realised, the overall behavior is dominated by $l_J(H)$, which gives a monotonic increase in $R_s(H)$, similar to that of $X_s(H)$.

Finally, the shape of $X_s(H)$ is very slightly affected by the change in $J_c(0)$, since the former is mostly sensitive to the form of $l_J(H)$ only, as can be shown by a simplified analysis of Eq.(13) and Eq.(15) in the limit of $\omega l_J < \rho_{Jf} \ll \rho_J$ (corresponding to the solid curve in Fig. 2 b). In the other limit considered here, $\rho_{Jf} \ll \omega l_J \ll \rho_J$, the inductive term $\omega l_J(H)$ still dominates the response, though $\rho_{Jf}(H)$ leads to a more linear shape of $X_s(H)$ [see Fig. 2 b, dotted line].

As was mentioned above, despite the fact that the present model is a multi-parameter theory, the physical arguments put rather strict limitations on the possible values of the parameters. Basically, it is the low-field junction resistivity, $\rho_{J0}$ and the critical current, $J_c(0)$ which produce the whole variety of the power dependences obtainable in the framework of this model, with the other parameters playing a minor role. For example, the parameter $\mu$ which adjusts the relative contribution of grains and weak links, does not change qualitatively the field dependences of $R_s$ and $X_s$, and is introduced here only to adjust the absolute change of $X_s$, $\Delta X_s$, in order to make it consistent with the experimental data.

The values of other parameters were taken from the literature to represent typical values obtained experimentally for high-quality YBaCuO thin films. For instance, values for $a$, $H_c$ and $J_c(0)$ can be found in Ref.[17], $\omega_{dJ}$ and $\eta_J$ in Refs.[17,27,28], $\rho_{J0}$ in Ref.[29], $\tau$ in Ref.[30]. Here, the value for Josephson vortex viscosity, $\eta_J$, in accordance with experimental findings of Ashkenazy et al.[28], were taken to be $\sim 2$ orders of magnitude lower than typical value for Abrikosov vortex viscosity, $\eta \sim 10^{-7}$ N·s/m$^2$. Tolerance of the simulated results to the parameter range involved was not too strict and, therefore, the main types of $Z_s(H)$ behavior can reliably be reproduced by slightly varying the input parameters.

It is an essential feature of the model that the junction resistivity both in the Meissner state, $\rho_J$ and flux flow regime, $\rho_{Jf}$ is considered field-dependent. The former is especially important for the non-shunted WL geometry to describe the correlated decrease of $R_s(H)$ and $X_s(H)$, whereas the latter is crucial for the shunted WL geometry. It is worth noting that under certain conditions the junction resistivity, which grows with the field, necessarily leads to the appearance of the maximum in $R_s(H)$, just as we observed for our films [see Fig. 2 a, dashed line]. A natural mechanism for this may be flux flow within the WLs, as suggested here, though a different mechanism cannot be not excluded. Here, the functional form of $\rho_{Jf}(H)$ is rather insignificant, whereas the general trend (an increase with the field) is crucial.

To summarize the various scenarii of our model, we should emphasize that the uncorrelated anomalies in Rs(Hrf) and Xs(Hrf) should be expected in HTS films containing shunted weak links. The uncorrelated behavior appears to be a characteristic feature of such type of the weak links. Here, the values of the junction critical current density, $J_c$ ($\sim 10^{10}$ A/m$^2$) and the quasiparticle resistivity, $\rho_{J0}$ ($\sim 10^{-6}$ $\Omega$·m) are rather typical for standard high-quality YBaCuO films.

As far as the correlated behavior is concerned, the HTS film should contain relatively low $J_c$ ($\sim 10^8$ A/m$^2$) and low $\rho_{J0}$ ($< 10^{-7}$ $\Omega$·m) weak links, which might be the regions of weakened superconducting properties displaying enhanced (compared to the bulk) normal carrier concentration and, as a consequence, lower $J_c$ and low $\rho_{J0}$ values. To the best of our knowledge, there is no evidence that such types of weak links naturally exist in HTS films, but it seems possible to fabricate an artificial junction with the aforementioned parameters (with the weak link material being a high-conductivity metal such as Cu, Ag or Au) to check out the idea of the correlated anomalous behavior. Here, $d$-wave nature of the superconducting electrodes of such a junction should make the anomalous effect quite noticeable (as it is seen in Fig. 4) due to (significantly) reduced scaling field, $H_g$ (or greater $b_\theta$) for the non-linear pair breaking in the $d$-wave superconductor.

Interestingly, recent dc transport experiments on artificial HTS junctions[31] have revealed anomalies in the tunneling conductance of the junctions. The zero bias conductance peak (ZBCP), which was previously theoretically predicted for an SNS junction made of $d_{x^2-y^2}$-wave superconductor[32], was experimentally observed in various HTS materials (YBaCuO, BSCCO and LSCO). This ZBCP is a result of the constructive interference between quasiparticles incident and reflected (due to Andreev reflection) from the grain-junction interface, and is possible only for a superconductor displaying zero energy bound states at the Fermi level (in order words, $d$-wave pairing). It would an interesting task for the future research to find out whether those transport anomalies have something in common with the microwave anomalies observed in HTS thin films.

In conclusion, the magnetic field, $H$, dependence of the surface impedance, $Z_s = R_s + jX_s$, of a Josephson weak link is simulated for two geometries; when the junction is shunted by another third grain, and for a non-shunted junction. In both cases, an anomalous behavior of $Z_s(H)$ is observed. In the case of the non-shunted junction, the correlated decrease of $R_s(H)$ and $X_s(H)$ is predicted which is due to a decreasing quasiparticle resistivity of the junction, $\rho_J$ being modulated by the field via the nonlinear pair breaking mechanism.

In the shunted junction geometry, non-monotonic behavior in $R_s(H)$ and $X_s(H)$ also arises, in this case being due to the junction resistivity, $\rho_{Jf}$ increasing with the field. The mechanism for such an increase is assumed to be flux flow in weak links.

It is worth noting that the Josephson junction like properties of HTS may not necessarily be due to the grain boundary effects, since they have also been seen in high-quality single crystals (see, e. g., Ref.[33]) with rather high $I_cR_n$ values. This might suggest that such properties are an intrinsic feature of the metal oxides owing to their complex pairing mechanism and extremely short coherence length, $\xi$.

The model presented here gives rather good qualitative (and in some cases even quantitative) description of the unusual features in $Z_s(H)$ of HTS reported so far.[5,4,7,8] However, we may not disregard some experimental observations (such as correlated decrease of $R_s(H)$ and $X_s(H)$, as well as an extremely high sensitivity of the anomalous $Z_s(H)$ behavior to relatively weak ($\sim$ 100 Oe) dc magnetic fields[8]) which might originate from the unconventional pairing mechanisms and complicated charge transfer mechanisms in HTS materials.

We would like to greatly acknowledge fruitful clarifying discussions with Prof. C. E. Gough, Prof. A. M. Portis, Dr. M. A. Hein, Prof. J. C. Gallop, Dr. M. Mahel, Dr. J. Halbritter, Prof. A. Ustinov, Dr. C. Muirhead, and Dr. A. P. Kharel. This work is supported by EPSRC grant No. RRHA04844.